\documentclass[sigconf]{acmart}

\usepackage{times}
\usepackage{moreverb}
\usepackage{graphicx}
\usepackage{mathrsfs}
\usepackage{bm}
\usepackage{url}
\usepackage{amsmath}
\usepackage{amssymb}

\usepackage{microtype}
\usepackage{graphicx}
\usepackage{subfigure}
\usepackage{balance}
\usepackage{booktabs} 
\usepackage{amssymb}
\usepackage{amsthm}
\usepackage{array}
\usepackage{graphicx}
\usepackage{clrscode}
\usepackage{subfigure}
\usepackage{multirow}
\usepackage{multicol}
\usepackage{float}
\usepackage{color}
\usepackage{xcolor}
\usepackage{amsopn}
\usepackage{mathrsfs}
\usepackage{mathtools}
\usepackage{amsmath}
\usepackage{booktabs}
\usepackage{arydshln}
\usepackage{hyperref}
\usepackage{blkarray}
\usepackage{enumerate}
\usepackage{courier}
\usepackage{mathrsfs}
\usepackage{rotating}
\usepackage{bm}
\usepackage{wrapfig}
\usepackage{subfigure}
\usepackage{array}
\usepackage{ragged2e}
\usepackage{hyperref}
\usepackage{amsmath}
\usepackage{threeparttable}    
\usepackage[ruled,linesnumbered]{algorithm2e}

\SetKwComment{comment}{ $triangleright$ \ }{}

\theoremstyle{definition}

\theoremstyle{theorem}

\theoremstyle{proof}

\theoremstyle{remark}

\AtBeginDocument{%
  \providecommand\BibTeX{{%
    \normalfont B\kern-0.5em{\scshape i\kern-0.25em b}\kern-0.8em\TeX}}}

\copyrightyear{2022}
\acmYear{2022}
\setcopyright{acmcopyright}

\begin{document}
\newcommand{\xu}[1]{{\color{red} xu: #1}}

\title{User Behavior Understanding In Real World Settings}
\author{Weiqi Shao$^{1,}$, Xu Chen$^{1,*}$, Jiashu Zhao$^{3}$,Long Xia$^{2}$, Dawei Yin$^{2}$}\thanks{$*$ Corresponding author}
\affiliation{\{$^1$Gaoling School of Artificial Intelligence\} Renmin University of China, Beijing 100872, China} 
\affiliation{$^2$Baidu Inc} 
\affiliation{$^3$Department of Physics and Computer Science, Wilfrid Laurier University} 
\affiliation{shaoweiqi@ruc.edu.cn, successcx@gmail.com, long.phil.xia@gmail.com , yindawei@acm.org }

\begin{abstract}
How to extract meaningful information in user historical behavior plays a crucial role in sequential recommendation. 
User's behavior sequence often contains multiple conceptually distinct items that belong to different item groups and the number of the item groups is changing over time.
It is necessary to learn a dynamic group of representations according the item groups in a user historical behavior.
However, current works only learns a predefined and fixed number representations which includes single representation methods and multi representations methods from the user context that could lead to suboptimal recommendation quality. 
In this paper we propose an Adasplit model that can automatically and adaptively generates a dynamic group of representations from the user behavior accordingly.
To be specific, 
AutoRep is composed of an informative representation construct (IRC) module and a dynamic representations construct (DRC) module.
The IRC module learns the overall sequential characteristics 
of user behavior with a bi-directional architecture transformer.
The DRC module dynamically allocate the item in the user behavior into different item groups and form a dynamic group of representations in a differentiable method. 
We formalize the hard allocation problem in the form of Markov Decision Process(MDP), and sample an action from Allocation Agent $\pi$
with a Group Controller Mechanism 
for each item to determine which item group it belongs to.
Such design improves the model’s recommendation performance. We evaluate the proposed model on five benchmark datasets. The results show that AutoRep outperforms representative baselines. Further ablation study has been conducted to deepen our understandings of AutoRep, including the proposed module IRC and DRC.
\end{abstract}
\maketitle

\section{Introduction}\label{introduction}
Recommender system has extensively permeated into people's daily life, ranging from the fields of e-commerce~\cite{li2017neural,liu2018stamp,wang2019modeling}, education~\cite{lin2018intelligent,saito2020learning} to the domains of health-caring~\cite{zhou2020cnn,gong2021smr,bhoi2020premier}, and entertainment~\cite{ayata2018emotion,subramaniyaswamy2019ontology,reddy2019content}.
In the past few years, many promising recommender models have been proposed, among which sequential recommendation is a type widely studied algorithm.
In sequential recommendation, the item that a user may interact with is estimated based on their historical behaviors.
Comparing with general recommender models like matrix factorization~\cite{he2017neural,xue2017deep,chen2020efficient}, sequential recommendation is advanced in the capability of capturing item correlations, which provides more informative signals to predict the next behavior.
In order to capture item correlations, recent years have witnessed quite a lot of effective sequential recommender models.
Essentially, these methods are built by introducing different assumptions on user successive behaviors.
For example, FPMC~\cite{rendle2010factorizing} believes that the current user behavior is only influenced by the most recent action, and thus regards user behaviors as a Markov chain. GRU4Rec~\cite{hidasi2015session} assumes that all the history behaviors are meaningful for the next item prediction, and therefore leverages recurrent neural network for summarizing all previous behaviors.

Despite effectiveness, existing algorithms usually assume user behaviors to be coherent, and leverage unified architecture to predict the next item.
However, this assumption may not hold in practice due to the diverse and complex user preferences.
As exampled in Figure
\hyperlink{intro}{1}, user A is an electronic enthusiast and also likes sports, thus she purchased both digital and sports products.
In this scenario, if the next item is a pair of running shoes, then the user preference on digital products can be less important or even bring noises.
For more clean and focused estimation, many studies~\cite{li2019multi,cen2020controllable,chen2021exploring,chen2021multi,tan2021sparse} propose to disentangle the history information into multi interest sub-sequences and form multi interest representation for users.
However, these models suffer from many significant limitations:
(1) To begin with, they mostly assume fixed number of interest sub-sequences, which contradicts with the diverse user personalities.
In Figure~\ref{intro}, the history information of user A contains two categories of products (\emph{i.e.}, digital and sports).
While the preference of user B spans over three domains including books, baby products and clothes.
Even for the same user, the number of interest may also vary with different contexts.
When a user is aimless, she may explore more diverse products.
While once her intent is determined, she may only interact with relevant and consistent items.
(2) Then, existing models fail to consider the evolving nature of user preference.
Also see the example in Figure
\hyperlink{intro}{1}, in the beginning, user A purchased a phone, which triggered the following interaction with the phone case.
Then her interest on digital products ended, and she began to care more about sports items, and purchased sport shirt, sweatpants and smart-watch.
If we see the smart-watch independently, it can be allocated to the digital products, while by considering the above user evolving preference, we know that the reason of purchasing smart-watch is more likely for sports, such as timing, counting calories, among others.
Previous models do not explicitly model such evolving preference, which may lead to inaccurate item allocation, and impact the final recommendation performance.

In order to solve the above problems, in this paper, we propose a novel sequential recommender model to \underline{\textbf{ada}}ptively di\underline{\textbf{s}}entangle user \underline{\textbf{p}}reference by considering its evo\underline{\textbf{l}}v\underline{\textbf{i}}ng na\underline{\textbf{t}}ure (called AdaSplit for short).
The key of our idea is to design a behavior allocator, which is able to automatically determine the number of sub-sequences based on user evolving preference.
To achieve this goal,
we regard the decomposition of user history behaviors as a Markov decision process (MDP).
At each step, the agent selects which sub-sequence the current item should be allocated to.
First,
we put the history behaviors into a bi-directional architecture transformer~\cite{vaswani2017attention} to learn the overall sequential characteristics 
of user behavior and equip the item with the global information which makes the next item allocation task efficient.
What's more, 
in order to adaptively determine the number of sub-sequences, we design a special action called ``creating a new sub-sequence''.
In the policy rolling out process, the sub-sequence number is gradually increased until reaching a stable value, where all the user diverse preferences have been explored.
The reward is associated with the similarity between the target item and candidate sub-sequences, and also designed to encourage orthogonality between the generated sub-sequences.
To avoid of generating too much sub-sequences, we introduce a curriculum reward, which adaptively penalizes the action of generating new sub-sequence.
In a summary, the main contributions of this paper can be concluded as follows:

$\bullet$ We proposed to build sequential recommender models by adaptively disentangling user preferences, which, to the best of our knowledge, is the first time in the recommendation domain.

$\bullet$ To achieve the above idea, we design a reinforcement learning (RL) model to allocate user behaviors and adaptively create new sub-sequences, which captures the evolving nature of user preference.

$\bullet$ 
We conduct extensive experiments on four datasets with several benchmarks to demonstrate the effectiveness of our model,
and for promoting this research direction,
we have released our project at https://no-one-xxx.github.io/Adasplit/.

\begin{figure}[t]
\centering
\setlength{\fboxrule}{0.pt}
\setlength{\fboxsep}{0.pt}
\fbox{
\includegraphics[width=0.80\linewidth]{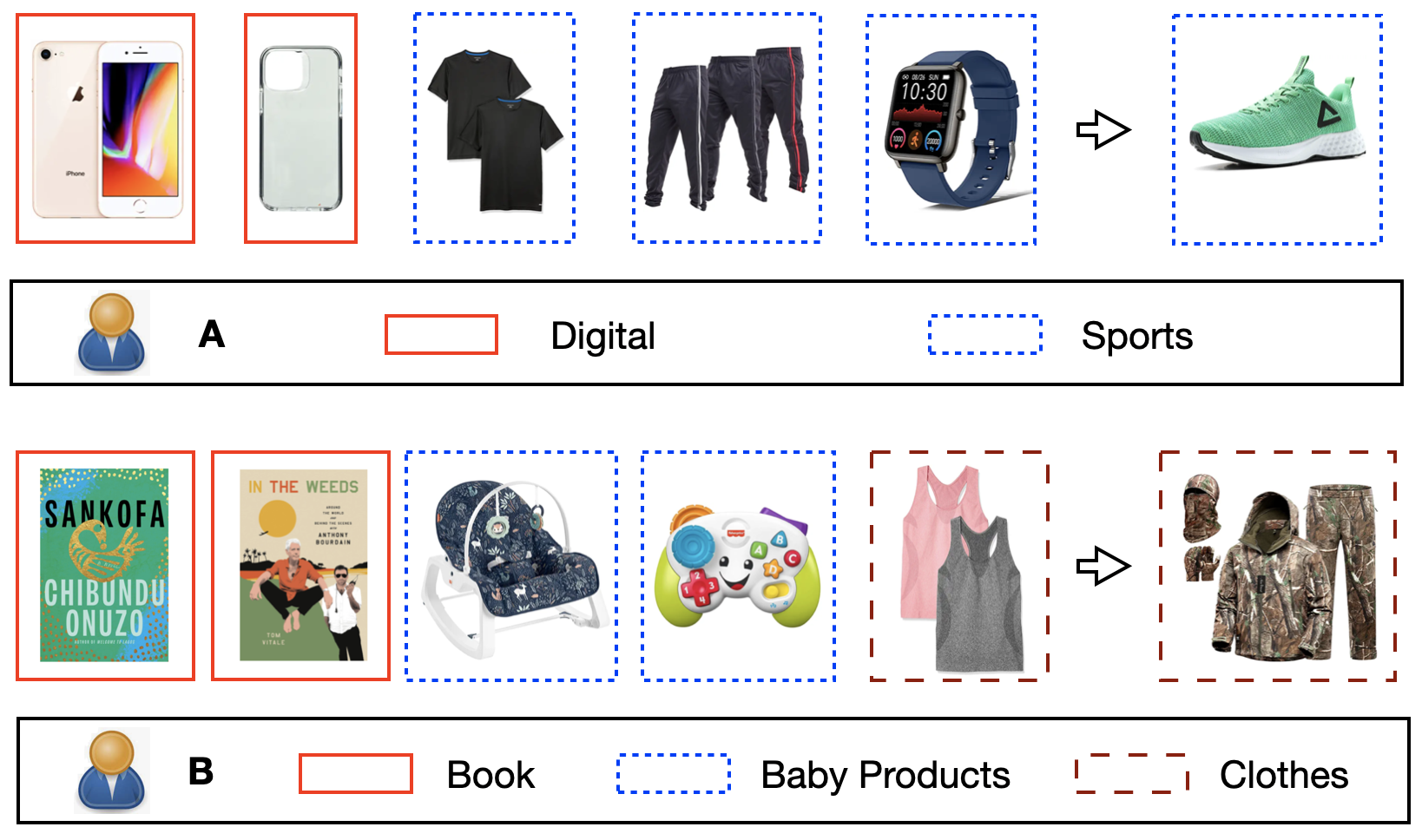}
}
\vspace*{-0.4cm}
\caption{
Illustration of user diverse preferences, where different users may have various personalities, and thus has different number of interests in their behavior sequence.
}
\label{intro}
\vspace*{-0.4cm}
\end{figure}

\section{Preliminary}

\subsection{Sequential Recommendation}
Sequential recommendation predicts the current user behavior by taking the history information into consideration.
Formally, we are provided with a user set $\mathcal{U}$ and an item set $\mathcal{V}$.
The interactions of each user $u\in \mathcal{U}$ are chronologically organized into a sequence $(v^1,v^2,...v^{l_u})$, which will be recurrently separated into training samples for model optimization.
For the sequence $(v^1,v^2,...v^{l_u})$, the generated samples are $\{[(u, v^1,v^2,...v^{j}),v^{j+1}]|j\in[1,l_u-1]\}$, where, in each sample, $(v^1,v^2,...v^{j})$ is the history information, and $v^{j+1}$ is the target item to be predicted.
We denote all the training samples for different users by $\mathcal{S} = \{[(u_i, v_i^1,v_i^2,...v_i^{l_i}),v_i^{l_i+1}]\}_{i=1}^N$.
Given $\{\mathcal{U},\mathcal{V},\mathcal{S}\}$, we have to learn a model $f$, which can accurately predict the next item that a user may interact with, given her history purchasing records. 
In the optimization process, sequential recommender models are usually learned based on the cross-entropy loss, that is:
{\setlength\abovedisplayskip{5pt}
\setlength\belowdisplayskip{5pt}
\hypertarget{gongsi 1}{}
\begin{eqnarray}\label{sin-seq}
\begin{aligned}
L_{seq} = -\sum_{i=1}^N \sum_{k=1}^{|\mathcal{V}|} y_k \log [f(u_i, v_i^1,v_i^2,...v_i^{l_i})]_k
\end{aligned}
\end{eqnarray}
}
where the output layer of $f(\cdot)$ is a softmax operation, and $[f(\cdot)]_k$ selects its $k$th element.
$y_k=1$ if $k = v_i^{l_i+1}$, otherwise $y_k=0$.
In the past few years, people have designed a lot of methods to implement $f$.
However, most of them assume user preference in the history information is coherent, which is less reasonable given the potentially diverse user personalities.

\subsection{Multi-interest Recommendation}
To solve the above problem, recent years have witnessed many multi-interests sequential recommendation algorithms.
Different from traditional methods, there is a multi interest extractor module $M$, which projects the history information into many multi interest representations.
Hopefully, each generated representation can exactly encode one type of user preference.
The next item is predicted based on the corresponding multi interest representations.                 
Formally, for the history information $(u_i,v_i^1,v_i^2,...v_i^{l_i})$.
Multi interest extractor $M$ first learn a fixed number multi interest representations(MIR) for the sequence behavior.
\hypertarget{gongsi 2}{}
{\setlength\abovedisplayskip{5pt}
\setlength\belowdisplayskip{5pt}
\begin{eqnarray}\label{mul-seq}
\begin{aligned}
MIR = M(u_i, v_i^1,v_i^2,...v_i^{l_i})
\end{aligned}
\end{eqnarray}
}

With MIR, 
current multi interest methods could been divided into two types in the prediction stage, where we call the divided method $d$: 
(1) ~\cite{cen2020controllable,chen2021exploring} use one of the interest representation in MIR for prediction.
They choose the target interest representation which gets the maximum inner product between the target item $v_i^{l_i+1}$.
(2) ~\cite{li2019multi,wang2019modeling,chen2021multi,tan2021sparse} use attention mechanisms which combines those interest representations with adaptive weights to represent the sequence behavior.
Both methods learn a behavior representation for the item prediction. 
In the training process, the learning objective is improved as:
\hypertarget{gongsi 3}{}
{\setlength\abovedisplayskip{5pt}
\setlength\belowdisplayskip{5pt}
\begin{eqnarray}\label{mul-seq}
\begin{aligned}
L_{mul} = -\sum_{i=1}^N \sum_{k=1}^{|\mathcal{V}|} y_k \log [d(MIR)]_k
\end{aligned}
\end{eqnarray}
}
where $y_{tk}=1$ if $k = v_i^{l_i+1}$ and $v_i^{l_i+1}$ falls into the $t$th user preference, otherwise $y_k=0$.
In this objective, the next item $v_i^{l_i+1}$ is predicted by considering all the multi interest representations or one of the multi interest representations.



\hypertarget{figure 2}{}
\begin{figure}[t]
\centering
\setlength{\fboxrule}{0.pt}
\setlength{\fboxsep}{0.pt}
\fbox{
\includegraphics[width=.9\linewidth]{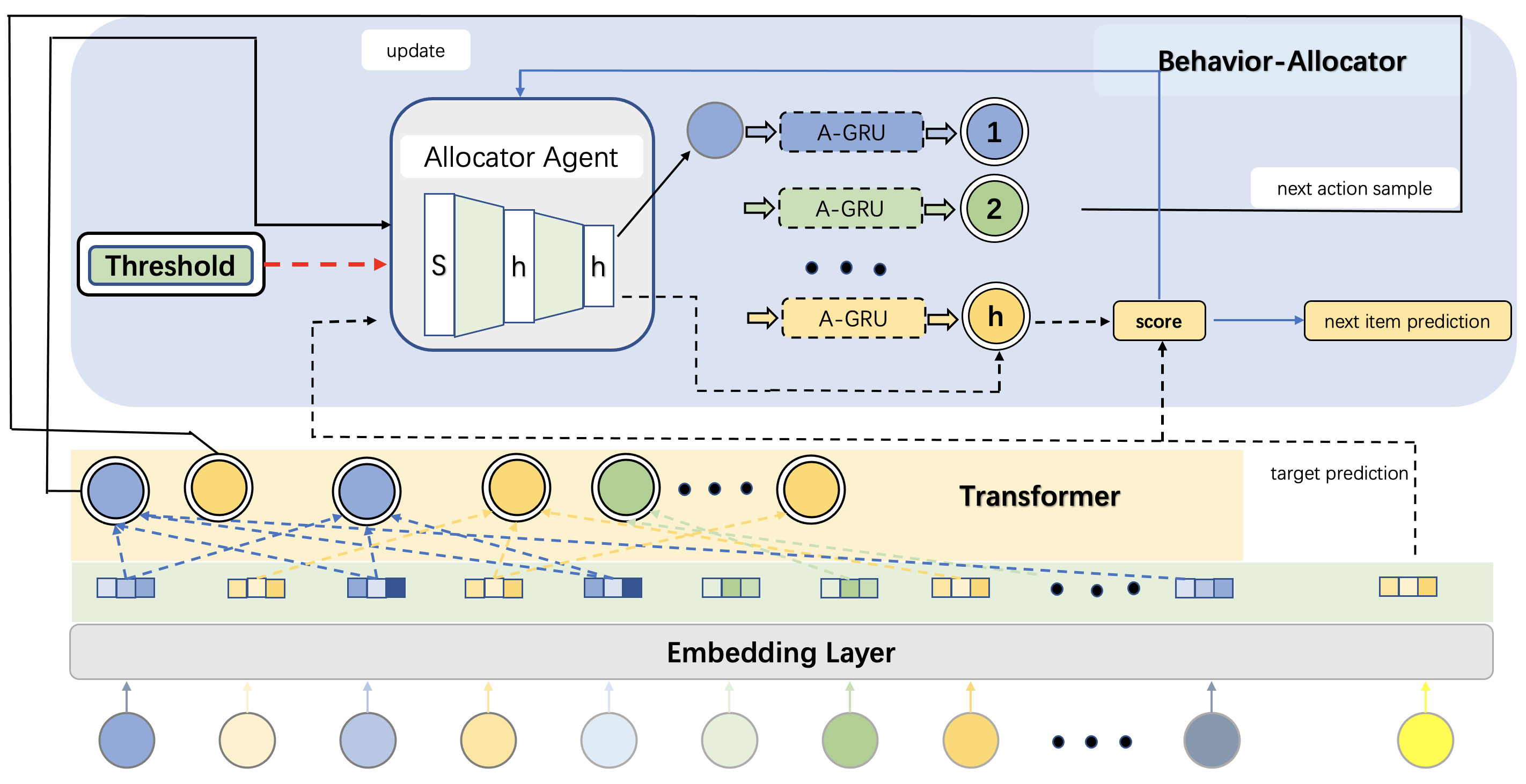}
}
\vspace*{-0.cm}
\caption{
The behavior allocator interactions in Adasplit.
}
\label{intro}
\vspace*{-0.3cm}
\end{figure}
\section{THE AdaSplit MODEL}
In our model, there are two major components: the first one is a reinforcement learning based allocator agent $\pi$, aiming to disentangle user preferences into different sub-sequences in 
\hyperlink{figure 2}{Figure 2}.
The other one is a sequential recommender model, which is leveraged to predict the next item, and generate the recommendation list.
In the following, we detail these components more in detail.
\subsection{Behavior Allocator}

The key of our model lies in how to allocate the items in history into different sub-sequences.
We first input the sequence behavior into a bi-directional architecture transformer, where
with the
bi-directional architecture transformer block, conceptually similar
item in the user behavior is fusing closer with the adaptive weights and the bi-directional architecture equips item  with the global information which
helps a lot in the item allocation task.

\textbf{Bi-Directional Transformer}
 First, we would omit the subscript i in user behavior. And we project user behavior $(u,v^1,v^2,...v^{l})$ to embedding vectors 
 $E = \{[(e_{u}, e^1,e^2,...e^{l}),e^{l+1}]\}_{i=1}^N$.
 Then, we incorporate a learnable position encoding matrix $P$ 
 to enhance the input representations. In this way, the input representations
 $E$ can be obtained by adding the two embedding matrices:
 $E=E+P$.
The attention layer calculates a weighted sum of all values, where the weight between query and key.
\hypertarget{gongsi 4}{}
{\setlength\abovedisplayskip{5pt}
\setlength\belowdisplayskip{5pt}
\begin{eqnarray}
\begin{aligned}
Attention(Q, K, V) = Softmax(\frac{QK^T}{\sqrt{d}})V
\end{aligned}
\end{eqnarray}
}

where the scale factor $\sqrt{d}$ is to avoid too large values of the inner product when the dimension is very high.
We take E as input, convert it to three matrices through linear projections, and feed them into an attention layer:
\hypertarget{gongsi 5}{}
{\setlength\abovedisplayskip{15pt}
\setlength\belowdisplayskip{5pt}
\begin{eqnarray}
\begin{aligned}
S &= Attention(EW^Q, EW^K, EW^V) \\\
S &= LayerNorm(S)
\end{aligned}
\end{eqnarray}
}

where the projections matrices $W^Q$ $W^k$ $W^v$ $\in$ $\mathcal{R}^{d*d}$.
The projections make the model more flexible.
And the we use LayerNorm to ensure the stability of the data feature distribution and accelerate the training process.
In order to enforce the model with non-linearity and to get more high-order interaction information, we apply a two-layer feed-forward network to all $S$.
\hypertarget{gongsi 6}{}
{\setlength\abovedisplayskip{15pt}
\setlength\belowdisplayskip{5pt}
\begin{eqnarray}
\begin{aligned}
v^{i} &= ReLU(S^iW^1+b^1)W^2+b^2 \\\
v^{i} &= LayerNorm(v^{i}+S^{i})
\end{aligned}
\end{eqnarray}
}

We add the original $s_{i}$ in \hyperlink{gongsi 6}{Eq.(6)}
which
could avoid network degradation and learn a more useful information in a deeper stack architecture deep model.

\textbf{Allocator Agent $\pi$}
We hope to adaptively determine the number of sub-sequences and also take the evolving nature of user preference into consideration.
To this end, we regard the separation of user behavior sequences as a Markov decision process, and design a reinforcement learning model for item allocation.

In particular, for a given sample, the agent goes through the history information, and at each step, it chooses a sub-sequence for the current item.
At step time = $T$, let all the generated sub-sequences be: $G^T = \{g_1^T,g_2^T,...,g_h^{T}\}$, 
where h is the current generated sub-sequences number.
What's more, we use sub-sequence representation $P^T = \{p_1^T,p_2^T,...,p_h^{T}\}$ to represent each generated sub-sequence and each
sub-sequence representation is initializing with the corresponding user embedding $e_u$.
The agent at step $T$ aims to assign the $T$th item $v^T$ in user's sequence behavior to a sub-sequence.
Given $G^T = \{g_1^T,g_2^T,...,g_h^{T}\}$ and the current state the
$S^T = \{s_1^T,s_2^T,...,s_h^{T}\}$ in 
\hyperlink{gongsi 13}{Eq.(13)} which is made up of the $S^T$ and $p_i^T$,
the allocator agent $\pi$ is implemented with the following policy network:
{
\setlength\abovedisplayskip{15pt}
\setlength\belowdisplayskip{5pt}
\hypertarget{gongsi 7}{}
\begin{eqnarray}
\begin{aligned}
\overline{r}_{j}^{T} = Sigmoid((ReLU(s^{T}_{i}W_{p1}+b_{p1})W_{p2}+b_{p2})W_{p3}+b_{p3}) \\
\pi(a^T = a|\bm{r}_1^T,\bm{r}_2^T,...\bm{r}_{h}^T) = [\text{SOFT-MAX}(\overline{r}_1^T,\overline{r}_2^T,...,\overline{r}_{h}^T)]_{a}
\end{aligned}
\end{eqnarray}
}

where $a^T$ is the action taken by the agent at step $T$, which means the new item $v^T$ belongs to the $a$ th sub-sequence $g_a^T$.
$W_{p1}$, $W_{p2}$ and $W_{p3}$ are adapting parameter for computing the similarity between $\bm{e}_{v^z}$ and $\bm{s}_j$.
$\text{SOFT-MAX}$ is the softmax operator.
As a result, $\pi(a^T = a|\bm{s}_1,\bm{s}_2,...\bm{s}_{h})$ is the probability that item $v^T$ is allocated into the $a$th sub-sequence.
By this equation, the policy network aims to select the sub-sequence in $g_a$, which is semantically more compatible with $v^T$.

However, since the number of sub-sequences is fixed, if an item is not coherent with any of existing sub-sequences, it has to be compromisely allocated into the sub-sequence which is not that similar. 
In such a scenario, a better solution is to adaptively determine the number of sub-sequences, so as to make sure that each sub-sequence is semantically coherent.
To achieve this idea, we introduce a special action called ``creating a new sub-sequence'', and the policy network is improved as:
\hypertarget{gongsi 8}{}
{\setlength\abovedisplayskip{10pt}
\setlength\belowdisplayskip{5pt}
\begin{eqnarray}\label{pi-imp}
\begin{aligned}
\pi(a_k = a|\bm{r}_1^T,\bm{r}_2^T,...\bm{r}_{h+1}^T) &= [\text{SOFT-MAX}(\overline{r}_1^T,\overline{r}_2^T,...,\overline{r}_{h}^T,\epsilon)]_{a}
\end{aligned}
\end{eqnarray}
}

where $\epsilon$ is a pre-defined hyper-parameter and we extend the vector before softmax with an additional dimension.
$a\in [1,h+1]$, and when $a = h +1$, we do not assign $v^T$ to any existing sub-sequences, but create a new sub-sequence, and regard $v^T$ as the first item in this new sub-sequence.
This equation encodes the following belief:
if $v^T$ is not close enough to any existing sub-sequences, that is, $\overline{g}_j < \epsilon, \forall j\in[1,h]$, then the policy network is more likely to choose action $h +1$, which creates a new sub-sequence.
However, if there are many sub-sequences, satisfying $\overline{r}_j > \epsilon$, then action $h +1$ may not be triggered.
As a special case, if $\overline{r}_j >> \epsilon, \forall j\in[1,h]$, then equation~(\ref{pi-imp}) is equal to equation~(\ref{pi}).

\textbf{Sequential recommender model}
Agent $\pi$ allocates the $v^T$ to the sub-sequence $g_a^T$, and we use the well-design sequential recommender model Attention-GRU to update the sub-sequence representation $p_a^T$, which is as follows:
\hypertarget{gongsi 9}{}
{\setlength\abovedisplayskip{15pt}
\setlength\belowdisplayskip{5pt}
\begin{eqnarray}
\begin{aligned}
p_a^{T+1} &= Attention-GRU(p_a^T,v^T)
\end{aligned}
\end{eqnarray}
}
And  the corresponding Attention-GRU is below:
\hypertarget{gongsi 10}{}
{
\setlength\abovedisplayskip{15pt}
\setlength\belowdisplayskip{5pt}
\begin{equation}
\begin{aligned}
z_i^{T+1} &= Sigmoid(W_z v^{T} + U_z p_i^T) \\
s_i^{T+1} &= Sigmoid(W_r v^{T} + U_r p_i^T) \\
h_i^{T+1} &= Tanh(W v^{T} + U(s_i^{T+1} * p_i^T)) \\
P_i^{T+1} &= z_i^{T+1} * p_i^{T} + (1-z_i^{T+1}) * h^{T+1}
\end{aligned}\label{1}
\end{equation}
}
where the output dimension of the $z_i^{T+1}$ and $s_i^{T+1}$ is 1.

\textbf{State Transition}
After an action is taken by the agent, the state is updated as follows:
{
\setlength\abovedisplayskip{5pt}
\setlength\belowdisplayskip{5pt}
\hypertarget{gongsi 11}{}
\begin{equation}
    G^T = 
    \begin{cases}
        \{(g_{1_1}^T,...,g_{1_{n_1}}^{T}),...(g_{h_1}^T,...,g_{h\_n_k}^{T},v^T),...\} \ ,& if \ a^T\in[1,h] \\\\
        \{(g_{1_1}^T,...,g_{1_{n_1}}^{T}),...(g_{h_1}^T,...,g_{h\_n_h}),(v^T)\} \ , & if \ a^T=h+1
    \end{cases}
\end{equation}
}


where if the action is in $[1,h]$, then the corresponding sub-sequence is extended with $v^T$.
If the action is ``creating a new sub-sequence'', then a new sub-sequence $(g_{h+1}^T)$ is added to the state, where the new sub-sequence representation is initializing with the corresponding user embedding $e_u$.


What's more, in reality, there are complex relationships between the user’s click sequence, like point level, union level with or without skip\cite{tang2018personalized}.
For accurately capturing those relationships,
we use a well-designed attention mechanism to define the state transition.
\hypertarget{gongsi 12}{}
{\setlength\abovedisplayskip{15pt}
\setlength\belowdisplayskip{5pt}
\begin{eqnarray}
\begin{aligned}
s^{T+1} &= concat(\sum_{i=1}^{h}z_{i}p_{i}^{T+1},v^{T+1})W^0 \\\
z_{i} &= \frac{\exp((p_{i}^T\cdot v^{T+1}))}{\sum_{i=1}^{h} \exp((p_{i}^T\cdot v^{T+1}))}
\end{aligned}
\end{eqnarray}
}

where $W^0$ is a $2d x d$ matrix and $(\cdot)$ represent the inner product. As for the sub-sequence state $s_i^{T+1}$ in 
\hyperlink{gongsi 7}{Eq.(7)}
is as follows:
\hypertarget{gongsi 13}{}
{\setlength\abovedisplayskip{10pt}
\setlength\belowdisplayskip{5pt}
\begin{eqnarray}\label{action_state}
\begin{aligned}
s^{T+1}_i = concat(s^{T+1}W_s,p_{T+1}W_p)
\end{aligned}
\end{eqnarray}
}

In our model, the reward is associated with the following aspects:

$\bullet$ \textbf{The allocation task loss.}
To begin with, we task of allocator agent $\pi$ is allocating item $v^T$ to the nearest sub-sequence.
Here we use the inner product between $v^T$ and its target sub-sequence representation $p_{a}^T$ as the accuracy of the "nearest".
And the reward of the allocation task is:
\hypertarget{gongsi 14}{}
{\setlength\abovedisplayskip{10pt}
\setlength\belowdisplayskip{5pt}
\begin{eqnarray}\label{seq}
\begin{aligned}
r^T_{loss} = \frac{\exp((p_{a}^T\cdot v^T))}{\sum_{i=1}^{h}
\exp((p_{i}^T\cdot v^T))}
\end{aligned}
\end{eqnarray}
}

where $h$ is the current generated sub-sequences number.
What's more,
the $\pi$ optimizes the relevance between $v^T$ and $p_{a}^T$, which is same as the the sequential recommender model optimize in 
\hyperlink{gongsi 20}{Eq.(20)}.
They are basically playing a collaborative game, where the common target is to lower the loss of the objective, that is, better fitting the training data.

$\bullet$ \textbf{The orthogonality between different sub-sequences.}
Ideally, each sub-sequence should encode just one type of user preference coherently, which is not leaked into the other ones.
Based on this property, each sub-sequence is representative enough, which facilitates more focused and explainable estimation.
By the design of our policy network $\pi$ in 
\hyperlink{gongsi 7}{Eq.(7)}
\hyperlink{gongsi 8}{(8)}
, we have tried to allocate each item into the sub-sequence which is most coherent with it.
Here, we introduce a reward to encourage that the information overlap between different sub-sequences is as small as possible, that is:
\hypertarget{gongsi 15}{}
{\setlength\abovedisplayskip{15pt}
\setlength\belowdisplayskip{5pt}
\begin{eqnarray}
\begin{aligned}
r^T_{ort}=-\frac{\sum_{i=1}^{h}\sum_{j=i+1}^{h}|p_i^t\cdot p_j^t|}{\frac{h*(h-1)}{2}}
\end{aligned}
\end{eqnarray}
}

where we try to make each pair of sub-sequence representations derived from equation
\hyperlink{gongsi 9}{Eq.(9)}
as orthogonal as possible.

$\bullet$ \textbf{The penalty on creating new sub-sequence.}
The special action ``creating new sub-sequence'' influences the final number of sub-sequences.
If there are too few sub-sequences, the user preference may not be well separated.
While if the number of sub-sequences is too large, the preference granularity can be too small, which may fail to capture the high-level connections among user behaviors, and thus may not generalize well on the complex and volatile recommendation environments.
In addition, if the number of sub-sequences is too large, then each sub-sequence may only contain very few items, which brings difficulties for learning each user preference sufficiently.

In order to tune the number of sub-sequences, we introduce a reward $r_{creat}$ to penalize the action of ``creating new sub-sequence'', that is:
{
\setlength\abovedisplayskip{10pt}
\setlength\belowdisplayskip{5pt}
\hypertarget{gongsi 16}{}
\begin{equation}
    r^T_{creat} = 
    \begin{cases}
        0 \ ,& if \ a^T\in[1,h] \\\\
        -\lambda^T \ , & a^T=h+1
    \end{cases}
\end{equation}
}

where the reward is a negative value after taking the special action ``creating new sub-sequence'', while for the other actions, the reward is 0.
By a larger $\lambda$, we can reduce the number of sub-sequences, while if $\lambda$ is small, then more sub-sequences can be generated.

In our model, we adjust $\lambda$ in a curriculum learning manner.
In the beginning, there is only one sub-sequence, and we do not hope to impose much constraint on ``creating new sub-sequence''.
At this time, $\lambda$ is a small value. 
While as the agent takes more actions, more sub-sequences are generated, thus we limit the number of sub-sequences by setting $\lambda$ as a larger value.
To realize the above idea, we explore $\lambda$ in the following set of functions when the agnet take the action "creating a new sub-sequence". And we use the $T_a$ as the times the action "creating a new sub-sequence" has shown up until now:
\hypertarget{gongsi 17}{}
\begin{equation}
    \lambda^{T+1} = 
    \begin{cases}
        \text{linear:}\quad &a_1*T_a\\\\
        \text{exponential:}\quad &b_1^{T_a}
    \end{cases}
\end{equation}

Where
$\{b_1, a_1\}$ are hyper-parameters, and we set $a_1>0, b_1>0$ to ensure that the function is monotonic as the agent takes more action "creating a new sub-sequence".

We sum the $r^T_{loss}$, $r^T_{ort}$ and $r^T_{creat}$ as the final reward $r^T$.
What's more, the current action value is only related to future rewards without considering previous rewards, we add the future rewards with a decay parameter $\lambda_d$.

\hypertarget{gongsi 18}{}
{\setlength\abovedisplayskip{5pt}
\setlength\belowdisplayskip{5pt}
\begin{eqnarray}\label{rl-loss}
\begin{aligned}
r^T = r^T_{loss} + r^T_{ort}*\lambda_o + r^T_{creat} \\
r^T=\sum_{c=T}^{l_i} \lambda_d^{c-T}r^T 
\end{aligned}
\end{eqnarray}
}

Where the $\lambda_o$ is the trade-off parameter to balance the importance of the orthogonal reward.
In practice, the discrete action is not differentiable, but we can optimize it with the log trick approximation, which is an unbiased estimation.
As a result, our final optimization target is:
\hypertarget{gongsi 19}{}
{\setlength\abovedisplayskip{5pt}
\setlength\belowdisplayskip{5pt}
\begin{eqnarray}\label{rl-loss}
\begin{aligned}
L_{rl} = -\sum_{i=1}^N \sum_{T=1}^{l_i}r^T*log(\pi(a^T|s^T))
\end{aligned}
\end{eqnarray}
}




\subsection{Model Optimization}
For the target item $v^{l+1}$, the behavior alloator $\pi$ generates its target sub-sequence $g_a^{l+1}$ and the target corresponding sub-sequence representation is $p_a^{l+1}$.
Then the sequential recommender model is optimized based on the following objective:
\hypertarget{gongsi 20}{}
{\setlength\abovedisplayskip{5pt}
\setlength\belowdisplayskip{5pt}
\begin{eqnarray}\label{seq}
\begin{aligned}
L_{seq}=-\log \frac{\exp((p_a^{l+1}\cdot v^{l+1}))}{\sum_{i=1}^{m}
\exp((p_a^{l+1}\cdot v_i))}
\end{aligned}
\end{eqnarray}
}

where the m is the number of the item.
Finally, we jointly train the allocator agent $\pi$ and 
sequential recommender with a trade-off parameter $\beta$:
\hypertarget{gongsi 21}{}
{\setlength\abovedisplayskip{5pt}
\setlength\belowdisplayskip{5pt}
\begin{eqnarray}
\begin{aligned}
\mathcal{L}=L_{seq}+\beta * L_{rl}
\end{aligned}
\end{eqnarray}
}

\subsection{Prediction}
When we do the prediction, allocator agent $\pi$ first 
allocate the item into the sub-sequences with the maximal probability in \hyperlink{gongsi 7}{Eq.(7)}
\hyperlink{gongsi 8}{(8)}
and then the state transition in \hyperlink{gongsi 11}{Eq. (11)(12)(13)}
, which can be written as follows:
\hypertarget{gongsi 22}{}
\begin{equation}
\begin{aligned}
a_{max}^T = argmax_{r}{\pi(a_k = a|\bm{r}_1^T,\bm{r}_2^T,...\bm{r}_{h+1}^T)}
\end{aligned}\label{1}
\end{equation}

We choose the action with the maximal probability in the prediction while we sample action according the action probability in the training process.
For each candidate item, allocator agent $\pi$ allocates it into a sub-sequence and calculates the inner product between the item representation and its target sub-sequence representation as the score.
We then rank all candidate items according to their scores 
and return the top-$N$ scores item as the final recommendations.



\subsection{Training Algorithm}
We summarize the complete learning algorithm of our framework
in Algorithm 1. 
The main task of allocator agent $\pi$ is allocating training sample $((u, v^1,v^2,...v^{l}),v^{l+1})$ into sub-sequences.
To begin with, the training sample is represented with a global information representation through a bi-directional architecture transformer in
\hyperlink{gongsi 5}{Eq.(5)}
\hyperlink{gongsi 6}{(6)}.
Then, the agent $\pi$ samples action for each item via its action prbability in 
\hyperlink{gongsi 7}{Eq.(7)}
\hyperlink{gongsi 8}{(8)} and allocates item 
into different sub-sequence $G$ and form sub-sequence representations $P$ in 
\hyperlink{gongsi 9}{Eq.(9)}
\hyperlink{gongsi 10}{(10)}.
And then state transition
in
        \hyperlink{gongsi 12}{(11)(12)}
        \hyperlink{gongsi 13}{(13)}
        and reward calculation in
\hyperlink{gongsi 14}{Eq.(14)}
        \hyperlink{gongsi 15}{(15)}
        \hyperlink{gongsi 16}{(16)}
for the item allocation. 
For the target item prediction, agent $\pi$ allocates the item into its target sub-sequence $g^{l+1}$ in 
\hyperlink{gongsi 7}{Eq.(7)}
\hyperlink{gongsi 8}{(8)}.
And optimization in the joint loss $\mathcal{L}$ in 
\hyperlink{gongsi 21}{Eq.(21)}.

\begin{algorithm}[t] 
\caption{Algorithm of AdaSplit}
\label{random-alg} 
Indicate the number of running times M.\\
Initialize all model parameters including $\pi$ and Attention-GRU.\\
\For{i in [0, M]}{
    Select a training sample $((u, v^1,v^2,...v^{l}),v^{l+1})$. \\
    Get global representation for $((u, v^1,v^2,...v^{l}),v^{l+1})$ in in
\hyperlink{gongsi 5}{Eq.(5)}
\hyperlink{gongsi 6}{(6)}.\\
    \underline{\textbf{Behavior Allocator:}} \\
    \For{T in [1, l]}{
        \underline{\textbf{Allocator agent $\pi$ allocates item $v^T$:}} \\
        1.Calculate state $s^T$ and $s^T_i$ in 
        \hyperlink{gongsi 12}{Eq.(12)(13)} \\
        2.Calculate $h$ generated sub-sequences state $s^T_i$ in 
        \hyperlink{gongsi 13}{Eq.(13)}\\
        3.With $s^T$ and $s^T_i$, $\pi$ derives action probability $\pi(a_k = a|\bm{r}_1^T,\bm{r}_2^T,...\bm{r}_{h+1}^T)$ in 
        \hyperlink{gongsi 7}{Eq.(7)}
        \hyperlink{gongsi 8}{(8)}\\
        4.Sample an action $a^T$ from $\pi(a_k = a|\bm{r}_1^T,\bm{r}_2^T,...\bm{r}_{h+1}^T)$\\
        5.Put the $v^T$ into the $a$ th sub-sequence $g_a^T$\\
        6.Update $a$ th sub-sequence representation $p^T_a$ via Attention-GRU in 
        \hyperlink{gongsi 9}{Eq.(9)}. \\
        7.Derive reward $r^T$ in 
        \hyperlink{gongsi 14}{Eq.(14)}
        \hyperlink{gongsi 15}{(15)}
        \hyperlink{gongsi 16}{(16)} \\
        8.Derive state transition $s^{T+1}$ in 
        \hyperlink{gongsi 12}{Eq.(11)(12)}
        \hyperlink{gongsi 13}{(13)} \\
        9.If $a^T$ is h+1: \\
        \qquad    Derive the $\lambda^{T+1}$ in 
        \hyperlink{gongsi 17}{Eq.(17)}. \\
    }
    With the generated sub-sequences $G^l$ and its representations $P^l$. \\
    \underline{\textbf{Model Optimization}}. \\
    Get the target sub-sequence $g_a^{l+1}$ for the $v^{l+1}$ in 
    \hyperlink{gongsi 7}{(7)}
    \hyperlink{gongsi 8}{(8)} \\
    $v^{l+1}$ and $g_a^{l+1}$ for model optimization: \\
    Calculate the RL loss $L_{rl}$ in 
    \hyperlink{gongsi 16}{(16)} \\
    Calculate the classification loss $L_{seq}$ in 
    \hyperlink{gongsi 19}{(19)}. \\
    Optimization the model with the joint loss $\mathcal{L}$ in 
    \hyperlink{gongsi 21}{Eq.(21)}
    
}
\end{algorithm}

\section{Experiments}
In this section, we will conduct experiments on four datasets to evaluate the effectiveness of Adasplit. 
We first briefly introduce the datasets and the state-of-the-art
methods, then we conduct experimental analysis on the proposed model and the benchmark models.
Specifically, we try to answer the following questions:

\indent
$\bullet$ \
How effective is the proposed method compared to other state-of-the-art baselines?
$\mathbf{Q1}$

\indent
$\bullet$ \
What are the effects of the bi-directional architecture transformer and
the sequential recommender model Attention-GRU in behavior allocator? 
$\mathbf{Q2}$

\indent
$\bullet$ \
How sensitive are the hyper-parameter the sequence length $t$ and the penalty parameter $\lambda$ in proposed model Adasplit?
$\mathbf{Q3}$


\subsection{Experimental Setup}
In this section, we introduce the details of the four experiment datasets,
evaluation metrics and comparing baselines in our experiments.

\textbf{Datasets.}
We perform experiments on four publicly available datasets,
including
$\mathbf{LastFM}$ \footnote{http://millionsongdataset.com/lastfm/}
a music records from Last.fm. And we only use the click behaviors.
$\mathbf{Garden, Baby, 
Beauty}$ \footnote{http://jmcauley.ucsd.edu/data/amazon/}
are composed of user purchasing behaviors from Amazon.
And the relative statistics information of the four datasets are shown in 
\hyperlink{table 1}{Table 1}.
For each datasets, we filter out items and users interacted less than five times.
And all datasets are taken Leave-one-out method in~\cite{kang2018self}
to split the datasets into training, validation and testing sets. 
Note that during testing, the input sequences contain training actions
and the validation actions for training the model.

\hypertarget{table 1}{}
\begin{table}[t]
\centering
\caption{\small{Statistics of the datasets.}}
\vspace{-0.3cm}
\small
\scalebox{.85}{
\begin{tabular}
{
p{2.4cm}<{\centering}|
p{1.2cm}<{\centering}|
p{1.2cm}<{\centering}|
p{1.8cm}<{\centering}|
p{1.2cm}<{\centering}}
\toprule[1pt]
       \textbf{Dataset}     &\textbf{\#User}   &\textbf{\#Item}
       &\textbf{\#Interaction}    &\textbf{Density}
      \\ \hline
{\textbf{Lastfm}}&1,860&2,824&71,355&1.36\\ 
{\textbf{Garden}}&1,687&963&13,272&0.82\\
{\textbf{Baby}}&19,446&7,051&160,792&0.12\\
{\textbf{Beauty}}&22,364&12,102&198,502&0.08b\\ 

\bottomrule[1pt]

\end{tabular}
}
\label{rec-dataset}
\vspace{-0.cm}
\end{table}



 


            

\begin{table*}[!t]
\hypertarget{table 2}{}
\caption{\small{Overall comparison between the baselines and our models Adasplit. 
The best results are highlighted with bold fold. All the numbers in the table are percentage 
numbers with '\%' omitted.
}}
\center
\small
\renewcommand\arraystretch{1.05}
\vspace{-0.4cm}
\setlength{\tabcolsep}{5.1pt}
\scalebox{.93}{

\begin{tabular}
{p{0.6cm}<{\centering}p{0.6cm}<{\centering}
p{0.6cm}<{\centering}p{0.6cm}<{\centering}|
p{0.6cm}<{\centering}p{0.6cm}<{\centering}
p{0.6cm}<{\centering}p{0.6cm}<{\centering}|
p{0.6cm}<{\centering}p{0.6cm}<{\centering}
p{0.6cm}<{\centering}p{0.6cm}<{\centering}|
p{0.6cm}<{\centering}p{0.6cm}<{\centering}
p{0.6cm}<{\centering}p{0.6cm}<{\centering}|
p{0.6cm}<{\centering}p{0.6cm}<{\centering}
p{0.6cm}<{\centering}p{0.6cm}<{\centering}
                      } \toprule[1pt]

\multicolumn{4}{c|}{} &
\multicolumn{4}{c|}{\textbf{LastFM}}&
\multicolumn{4}{c|}{\textbf{Garden}} &
\multicolumn{4}{c|}{\textbf{Baby}}&
\multicolumn{4}{c}{\textbf{Beauty}}\\ 
 
\multicolumn{4}{c|}{} & 
\multicolumn{4}{c|}{Metric@5 \qquad Metric@10} &
\multicolumn{4}{c|}{Metric@5 \qquad Metric@10} &
\multicolumn{4}{c|}{Metric@5 \qquad Metric@10} &
\multicolumn{4}{c}{Metric@5 \qquad Metric@10} \\ \hline

\multicolumn{4}{c|}{\textbf{Single Representation}} & 
\textbf{NDCG} & \textbf{MRR} & \textbf{NDCG} & \textbf{MRR} &
\textbf{NDCG} & \textbf{MRR} & \textbf{NDCG} & \textbf{MRR} &
\textbf{NDCG} & \textbf{MRR} & \textbf{NDCG} & \textbf{MRR} &
\textbf{NDCG} & \textbf{MRR} & \textbf{NDCG} & \textbf{MRR} \\ \hline

\multicolumn{4}{c|}{\textbf{BPR}} & 
0.31 & 0.23 & 0.41 & 0.27 &
1.72 & 1.37 & 2.25 & 1.59 &
0.75 & 0.61 & 0.98 & 0.70 &
1.69 & 1.37 & 2.21 & 1.59 \\  

\multicolumn{4}{c|}{\textbf{NeuMF}} & 
0.31 & 0.23 & 0.44 & 0.2 &
1.69 & 1.35 & 2.22 & 1.56 &
0.66 & 0.54 & 0.86 & 0.62 &
1.63 & 1.32 & 2.07 & 1.50 \\ 

\multicolumn{4}{c|}{\textbf{GRU4Rec}} & 
5.74 & 5.01 & 7.12 & 5.59 &
4.62 & 3.88 & 5.89 & 4.39 &
1.68 & 1.38 & 2.18 & 1.58 &
3.02 & 2.59 & 3.86 & 2.86 \\  

\multicolumn{4}{c|}{\textbf{STAMP}} & 
5.42 & 4.81 & 6.02 & 5.06 &
4.90 & 4.22 & 6.32 & 4.81 &
1.66 & 1.41 & 2.10 & 1.59 &
3.44 & 3.04 & 3.98 & 3.26 \\ 

\multicolumn{4}{c|}{\textbf{NARM}} & 
6.15 & 5.14 & 7.68 & 5.76 &
5.21 & 4.31 & 6.64 & 4.89 &
1.67 & 1.40 & 2.14 & 1.59 &
3.14 & 2.68 & 3.86 & 2.97 \\ 

\multicolumn{4}{c|}{\textbf{SASRec}} & 
7.37 & 6.42 & 8.61 & 6.91 &
5.33 & 4.30 & 7.34 & 5.12 &
1.68 & 1.32 & 2.21 & 1.54 &
3.63 & 2.92 & 4.54 & 3.29 \\  
\hline

\multicolumn{4}{c|}{\textbf{Multi Representations}} & 
\textbf{NDCG} & \textbf{MRR} & \textbf{NDCG} & \textbf{MRR} &
\textbf{NDCG} & \textbf{MRR} & \textbf{NDCG} & \textbf{MRR} &
\textbf{NDCG} & \textbf{MRR} & \textbf{NDCG} & \textbf{MRR} &
\textbf{NDCG} & \textbf{MRR} & \textbf{NDCG} & \textbf{MRR} 
\\ \hline

\multicolumn{4}{c|}{\textbf{MCPRN}} & 
5.92 & 5.12 & 6.68 & 5.44 &
5.46 & 4.62 & 6.82 & 5.17 &
1.90 & 1.64 & 2.37 & 1.83 &
3.69 & 3.24 & 4.29 & 3.48 \\ 

\multicolumn{4}{c|}{\textbf{SASRec-2}} & 
7.33 & 6.43 & 8.63 & 6.97 &
5.22 & 4.30 & 6.77 & 4.52 &
1.69 & 1.32 & 2.21 & 1.54 &
3.40 & 2.78 & 4.16 & 3.09 \\

\hline

\multicolumn{4}{c|}{\textbf{Dynamic Representations}} & 
\textbf{NDCG} & \textbf{MRR} & \textbf{NDCG} & \textbf{MRR} &
\textbf{NDCG} & \textbf{MRR} & \textbf{NDCG} & \textbf{MRR} &
\textbf{NDCG} & \textbf{MRR} & \textbf{NDCG} & \textbf{MRR} &
\textbf{NDCG} & \textbf{MRR} & \textbf{NDCG} & \textbf{MRR}
\\ \hline

\multicolumn{4}{c|}{\textbf{Adasplit}} & 
\textbf{7.88} & \textbf{7.05} & \textbf{8.72} & \textbf{7.39} &
\textbf{6.43} & \textbf{5.54} & \textbf{7.59} & \textbf{6.00} &
\textbf{2.00} & \textbf{1.73} & \textbf{2.44} & \textbf{1.90} &
\textbf{4.23} & \textbf{3.74} & \textbf{4.85} & \textbf{3.99} \\ 
\bottomrule[1pt]

\end{tabular}
}
            
\label{tab:ab-result}   
\vspace{-0.3cm}
\end{table*}



\textbf{Baeslines.}
We compare our proposed model Adasplit with the following state-of-the-art sequential recommendation baselines.

\begin{itemize}
\item \textbf{Single representation models:}
\textbf{BPR}~\cite{rendle2012bpr} is a famous recommendation algorithm for capturing user implicit feedback.
\textbf{NeuMF}~\cite{he2017neural} is a classic work which leverages the neural networks for capturing the user preference.
\textbf{GRU4Rec}
~\cite{hidasi2015session} is a pioneering work which first leverages GRU to model user behavior sequences for prediction. 
\textbf{STAMP}~\cite{liu2018stamp}
is a recently proposed neural network-based method capturing sequential pattern by emphasizing the user short preference.
\textbf{NARM}~\cite{li2017neural}
is a recently proposed neural attention based method.
\textbf{SASRec}~\cite{tang2018personalized}
is a well-known attention based sequential recommendation.
And the number of head in experiments is 1.
\item \textbf{Multi representations models:}
\textbf{MCPRN}~\cite{wang2019modeling} is a recent representative work for extracting multiple interests.
\textbf{SASRec-2}~\cite{tang2018personalized}
has two heads in the representations construction, which is a multi interest recommendation method~\cite{tan2021sparse}.
\end{itemize}

\textbf{Parameter Configuration.}
For a fair comparison, all baseline methods are implemented in Pytorch and optimized with Adam optimizer.
Specifically, the learning rate and batch size are tuned in
the ranges of [0.01,0.001,0.0001] and [32,64,128,256]. For our method, it has three crucial hyper-parameters: the "creating a new sub-sequence" parameter $\epsilon$ is tuned in the ranges of [0.2,0.3,0.4,0.5,0.6,
0.7,0.8],
and trade-off parameter $\lambda_0$ and $\beta$ are tuned in the ranges of [1,0.1,
0.01,0.001].
And the penalty parameter $\lambda$ is tuned in the ranges of [0.9,1,1.1,1.2,1.3].
What's more, we have released our project at
https://no-one-xxx.github.io/Adasplit/

\textbf{Evaluation Metrics.}
We use two commonly used evaluation criteria: Mean Reciprocal Rank (MRR) and Normalized Discounted Cumulative Gain (NDCG) to evaluate the performance of our model. 

\hypertarget{figure 4}{}
\begin{figure}[t]
\centering
\setlength{\fboxrule}{0.pt}
\setlength{\fboxsep}{0.pt}
\fbox{
\includegraphics[width=0.9\linewidth]{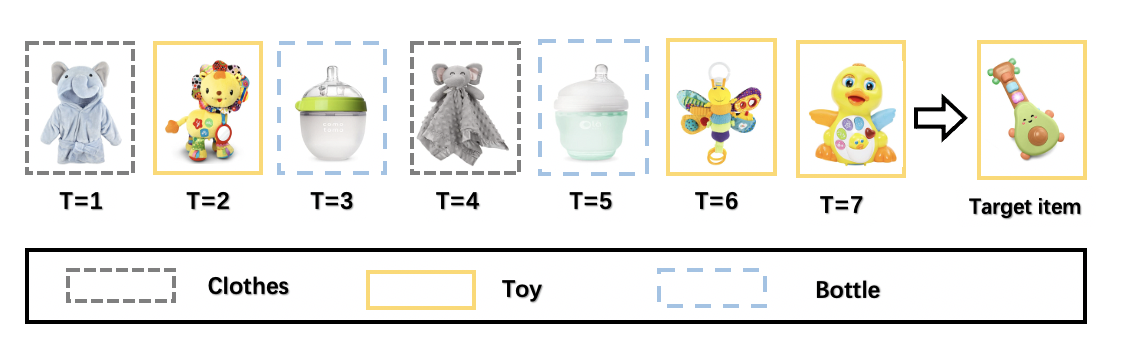}
}
\vspace*{-0.cm}
\caption{
Case study.
    The left before t=7 is the user behavior and the task to predict the target item at t=8.
    And the picture of each movie is downloaded from https://www.amazon.com.
}

\label{intro}
\vspace*{-0.3cm}
\end{figure}

\subsection{Overall performance (Q1)}
\hyperlink{table 2}{Table 2}
 summarizes the performance of Adasplit and baselines. Obviously, Adasplit achieves better performance to other the baselines on all the evaluation metrics. 
 
 
 What's more,
 in the datasets Amazon, compare single representation methods with multi representation methods and Adasplit,
 it is obvious that recommendation with multiple presentations (MCPRN) and with dynamic representation Adasplit for a user click sequence perform generally better performance than those with single representation (Caser, GRU4Rec, BERT4Rec ...).
 The user's always click multiply concept different items in those datasets which gives the evidence that only single representation for next item prediction can't handle the complex situations. However, in other datasets like LastFM, the multi representations method (MCPRN) doesn't achieve excellent results,
 which indicates that users in Lastfm datases show a focus click patterns and single representation would achieve better results.
 And our model Adasplit could learns a dyanmic group of representations according the user sequence behavior. In other words, Adasplit could learn more representations when the sequence behavior is complex and learn less representations when the sequence behavior is simple. And it shows its adaptability when it faces various datasets.
 
 
 The improvement of Adasplit over the multi representations methods (MCPRN, SASRec-2) and single representation methods (BPR, 
 NARM...) shows that the dynamic representations exploration serves as a better information extractor than fixed and predefined number representations. This can be attributed to two points: 
 1) Adasplit adjusts the sub-sequence representations construction according to data characteristics which could take the advantages of both the single representation methods and multi representations methods. 
 2) Adasplit considers the user's evolving preference in the user representations construction which gets high adaptability.
 
 In our model, a major novelty is that we want to allocate the item into different sub-sequences and learn a dynamic group of representations to represent the user. 
 To obtain a better understanding why Adasplit performs better than other models, we further construct a case study on Baby dataset. Specifically, we present a snapshot of the interaction sequence from a sampled user, which contains eight items.
 The results show in 
 \hyperlink{figure 4}{Figure 3},
 and we use different colors to represent different sub-sequences that the allocator agent $\pi$ generates.
 There are three different sub-sequences:
 clothes, 
 toy and bottle,
 which are fully in line with the facts in the user sequence behavior.
 As for the next item prediction, the target item is a toy guitar and it is accurately allocated to the toy sub-sequence exactly by the allocator agent $\pi$.
 
\hypertarget{figure 5}{}
\begin{figure}[t]
\centering
\setlength{\fboxrule}{0.pt}
\setlength{\fboxsep}{0.pt}
\fbox{
\includegraphics[width=.95\linewidth]{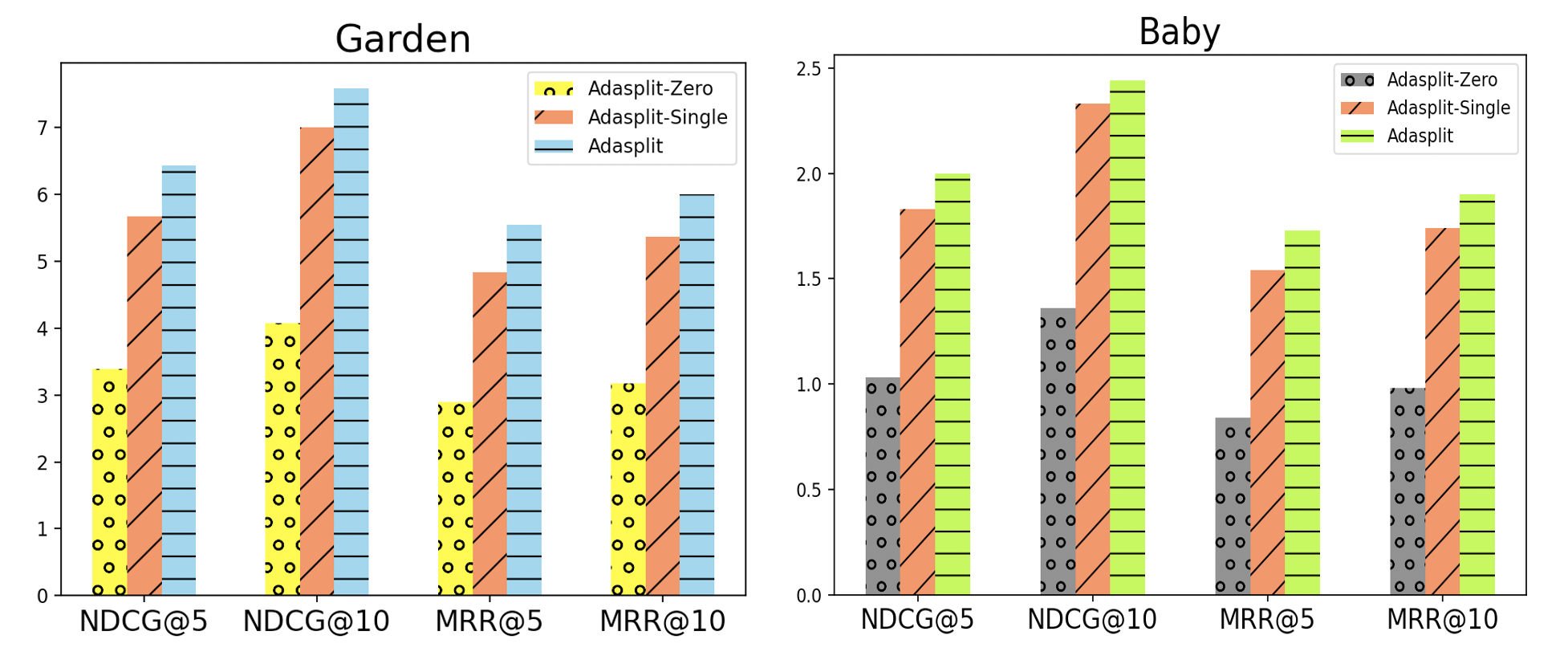}
}
\vspace*{-0.cm}
\caption{
Ablation study in bi-directional architecture transformer. 
Performance comparison of Adasplit, its variant Adasplit-Single and Adasplit-Zero.
}
\label{intro}
\vspace*{-0.3cm}
\end{figure}

\subsection{Ablation study (Q2)}

In behavior allocator, we use the bi-directional architecture transformer to equip item representations with global information for the next allocation task.
In order to explore the effectiveness of the bi-directional architecture transformer,
we change the bi-directional architecture transformer to left-to-right directional architecture transformer(Adasplit-Single) and use the original item sequence representation without any operation for the allocation task (Adasplit-Zero).
The item in Adasplit-Single could obtain the information form the items on its left while item in Adasplit-Zero knows nothing about the items around itself.
From the results reports in \hyperlink{figure 5}{Figure 4},
we can find that Adasplit-Zero achieves the worse performance compare with Adasplit-Single,
which indicates that the information between items is important in allocation process.
And the Adasplit achieves the better performance than the Adasplit-Single gives the evidence that the effectiveness of the bi-directional architecture transformer for extracting the global information.
In other words, two-way information is more useful than only one-way information in item allocation.

In behavior allocator, the sequential recommender model Attention-GRU formalize the sub-sequence representation.
To validate the effectiveness of the recommender model Attention-GRU,
we change the Attention-GRU to two other methods, LSTM and AveragePooling.
And We call the two variants Adasplit-LSTM, Adasplit-AveragePooling. 
From the results reports in \hyperlink{table 3}{Table 3},
Adasplit achieves a better performance compare with other variants, which validates the effectiveness of the sequential recommender model Attention-GRU in behavior allocator. 
And we also find Adasplit-LSTM achieve better performance to Adasplit-AveragePooling in the \hyperlink{table 3}{Table 3} which gives the evidence that the necessity of the sequential pattern to capture the evolving user preference.


\begin{table}[H]
\hypertarget{table 3}{}
\centering
\caption{\small{Ablation study in Attention-GRU. 
Performance comparison of Adasplit, its variants Adasplit-LSTM and Adasplit-AveragePooling.}}
\vspace{-0.3cm}
\small
\scalebox{.75}{
\begin{tabular}
{p{0.6cm}<{\centering}p{0.6cm}<{\centering}
p{0.6cm}<{\centering}p{0.6cm}<{\centering}|
p{0.6cm}<{\centering}p{0.6cm}<{\centering}
p{0.6cm}<{\centering}p{0.6cm}<{\centering}|
p{0.6cm}<{\centering}p{0.6cm}<{\centering}
p{0.6cm}<{\centering}p{0.6cm}<{\centering}
                      } \toprule[1pt]

\multicolumn{4}{c|}{} &
\multicolumn{4}{c|}{\textbf{Garden}} &
\multicolumn{4}{c}{\textbf{Baby}}\\ 
 
\multicolumn{4}{c|}{} & 
\multicolumn{4}{c|}{Metric@5 \qquad Metric@10} &
\multicolumn{4}{c}{Metric@5 \qquad Metric@10} \\ \hline

\multicolumn{4}{c|}{} & 
\textbf{NDCG} & \textbf{MRR} & \textbf{NDCG} & \textbf{MRR} &
\textbf{NDCG} & \textbf{MRR} & \textbf{NDCG} & \textbf{MRR} \\ \hline

\multicolumn{4}{c|}{\textbf{Adasplit}} & 
6.43 & 5.54 & 7.59 & 6.00 &
2.00 & 1.73 & 2.44 & 1.90 \\  

\multicolumn{4}{c|}{\textbf{Adasplit-LSTM}} & 
6.14 & 5.26 & 7.54 & 5.83 &
1.59 & 1.31 & 2.09 & 1.52 \\

\multicolumn{4}{c|}{\textbf{Adasplit-AveragePooling}} & 
4.47 & 3.41 & 6.63 & 4.28 &
1.56 & 1.23 & 2.05 & 1.42 \\  

\bottomrule[1pt]

\end{tabular}  
}
\label{rec-dataset}
\end{table}

\subsection{Hyperparameter study (Q3)}
As we all know, more item in user sequence behavior, more sub-sequences may occur.
Thus, the sequence length $l$ in the Adasplit is important,
and we do experiments to investigate the sensitivity of
the sequence length $l$ in Adasplit. 
\hyperlink{figure 6}{Figure 5}
reports the performance of our model in the metrics of MRR and NDCG in Garden and Baby datasets. 
In particular,
We keep the other parameters in the model consistent with the Q1 settings. 
From the figure, we
can observe that Adasplit obtains the best performance of MRR and NDCG when $t$ equals 10. 
The result 
increases with the increase of
the sequence length $l$. After it comes to a peak, it begins to decrease.

\hypertarget{figure 6}{}
\begin{figure}[t]
\centering
\setlength{\fboxrule}{0.pt}
\setlength{\fboxsep}{0.pt}
\fbox{
\includegraphics[width=.95\linewidth]{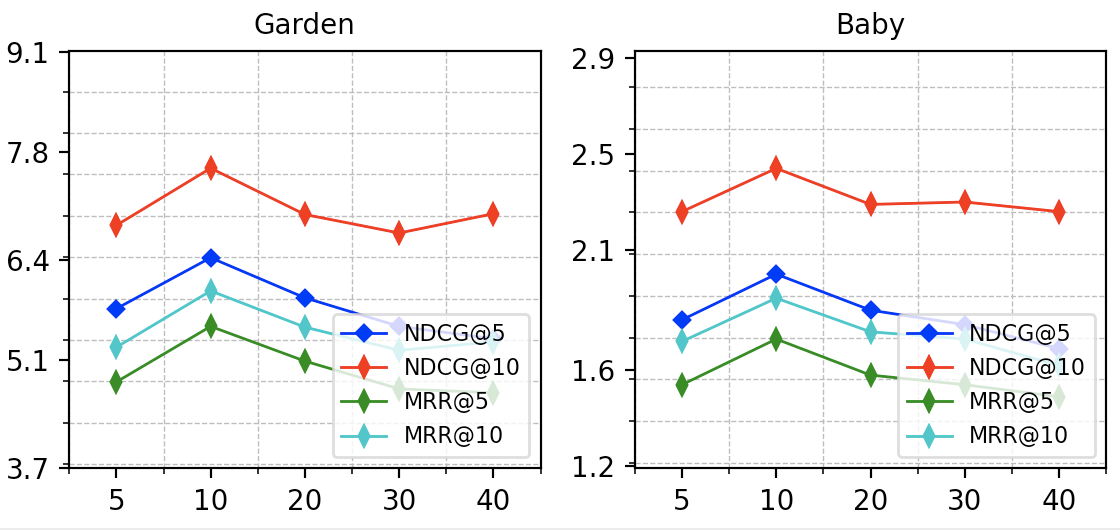}
}
\vspace*{-0.cm}
\caption{
Hyperparameter study, where the 
horizontal coordinates is the sequence length $l$ from 5 to 40.
}
\label{intro}
\vspace*{-0.3cm}
\end{figure}

In overall performance, Q1, we use the exponential growth in 
\hyperlink{gongsi 17}{Eq.(17)} for penalty $\lambda$.
Here we use the linear growth for the $\lambda$ in behavior allocator.
And we call it Adasplit-Linear.
In order to valid the effectiveness of the penalty operation and the increasing penalty $\lambda$ method on the action "creating a new sub-sequence".
We add another two variants, Adasplit-None, where we remove the penalty operation and Adasplit-Keep where we don't change the penalty parameter $\lambda$.
In \hyperlink{table 4}{Table 4},
obviously, Adasplit-None achieves the worst performance among the four models, which gives the evidence that the effectiveness of the penalty operation.
And Adasplit-Linear achieve better performance than Adasplit-Keep shows the effectiveness of the increasing penalty $\lambda$ method.
Last but not the least, Adasplit-Linear achieves worse performance compare with Adasplit, which gives us the evidence that the stronger increasing method may bring a better performance.

\begin{table}[H]
\hypertarget{table 4}{}
\centering
\caption{\small{Hyperparameter study, the performance of Adasplit and its variant Adasplit-Linear, Adasplit-Keep and Adasplit-None.}}
\vspace{-0.3cm}
\small
\scalebox{.85}{
\begin{tabular}
{p{0.6cm}<{\centering}p{0.6cm}<{\centering}
p{0.6cm}<{\centering}p{0.6cm}<{\centering}|
p{0.6cm}<{\centering}p{0.6cm}<{\centering}
p{0.6cm}<{\centering}p{0.6cm}<{\centering}|
p{0.6cm}<{\centering}p{0.6cm}<{\centering}
p{0.6cm}<{\centering}p{0.6cm}<{\centering}
                      } \toprule[1pt]

\multicolumn{4}{c|}{} &
\multicolumn{4}{c|}{\textbf{Garden}} &
\multicolumn{4}{c}{\textbf{Baby}}\\ 
 
\multicolumn{4}{c|}{} & 
\multicolumn{4}{c|}{Metric@5 \qquad Metric@10} &
\multicolumn{4}{c}{Metric@5 \qquad Metric@10} \\ \hline

\multicolumn{4}{c|}{\textbf{Model}} & 
\textbf{NDCG} & \textbf{MRR} & \textbf{NDCG} & \textbf{MRR} &
\textbf{NDCG} & \textbf{MRR} & \textbf{NDCG} & \textbf{MRR} \\ \hline

\multicolumn{4}{c|}{\textbf{Adasplit}} & 
6.43 & 5.54 & 7.59 & 6.00 &
2.00 & 1.73 & 2.44 & 1.90 \\  

\multicolumn{4}{c|}{\textbf{Adasplit-Linear}} & 
6.11 & 5.26 & 7.26 & 5.73 &
1.81 & 1.54 & 2.26 & 1.72 \\  

\multicolumn{4}{c|}{\textbf{Adasplit-Keep}} & 
5.99 & 5.08 & 6.88 & 5.43 &
1.67 & 1.34 & 1.99 & 1.46 \\ 

\multicolumn{4}{c|}{\textbf{Adasplit-None}} & 
4.32 & 3.86 & 5,87 & 4.63 &
1.25 & 1.08 & 1.65 & 1.15 \\ 

\bottomrule[1pt]

\end{tabular}

}
\label{rec-dataset}
\end{table}


\section{Related work}
In this section, we will briefly introduce the related works to our study, which includes sequential recommendation and multi interest recommendation

\subsection{Sequential Recommendation}
The main purpose of sequential recommendation is to discover the underlying
patterns of the user sequential behaviors.
Rendle et al~\cite{rendle2010factorizing} integrates matrix factorization and the sequential pattern of Markov Chains for prediction, and later Wang et al~\cite{wang2015learning} simultaneously consider the sequence behaviors and user preferences
with a hierarchical representation model.
Though those methods make progress in recommendation, these methods only model the local sequential patterns between recent clicked item ~\cite{yu2016dynamic}.
To model the longer sequential behaviors in user behavior, Hidasi et al~\cite{hidasi2015session} first adopted recurrent neural network to model the long sequence pattern. Then, Li et al~\cite{li2017neural} not only consider the sequence pattern in
the sequence, but also explore the user’s main purpose through the
attention mechanism.
And Ma et al~\cite{ma2019hierarchical} takes Hierarchical Gating Networks to capture both the long-term and short-term user interests.
 Chen et al~\cite{chen2018sequential} use Memory Network for exploring the sequential pattern in user behavior.
What's more, Tang et al~\cite{tang2018personalized} and Yuan et al\cite{yuan2019simple} embed the user historical sequence behavior 
into an “image”  and learn sequential
patterns as local features of the image with Convolutional Neural Network.
Later, Kang et al~\cite{kang2018self} considers the importance between different items in the click sequence and fuses the item representation with adaptive weights, which achieves great progress in many real datasets.
Wu et al~\cite{wu2019session}
and Huang et al~\cite{huang2021graph} use graph to learn the complex transitions between items and improve the recommendation performance.

\subsection{Multi Interest Recommendation}
Multi interest recommendation with multi interest representations have greater expressive power especially when the user shows a wide range of intends.
Liu et al~\cite{liu2019hi}
proposes a new user representation
model to comprehensively extract user sequence behavior
into multiple vectors.
And Xiao et al~\cite{xiao2020deep} explores user diverse interests with a multi-head architecture self-attentive, where the number of the heads is the number of the representations.
Wang et al~\cite{wang2019modeling}
takes an effective mixture-channel purpose
routing networks to detect the purposes of each item
and assigns items into the corresponding channels to get the multi interest representations.
Chen et al~\cite{chen2021exploring} thinks that the time interval information is meaningful in extracting the useful information and designs a novel time graph to get the time information in multi representations construction and Tan et al~\cite{tan2021sparse} infers a sparse set of concepts for each user from the large concepts to generate user multi interest representations. 
Chen et al~\cite{chen2021multi} and Cen et al~\cite{cen2020controllable} use capsule routing and self-attention as multi interest representations extractor and the proposed models improve the diversity for the next-item prediction.
Though those methods have achieved good performance in recommendation, none of them consider the diverse interests between users and the evolving user pattern.

\section{Conclusion}
In this paper, we proposed a novel model called Adasplit, to improve the recommendation performance by learning a dynamic group representations from the user's sequence behavior.
Adasplit can adaptively and automatically allocates items into different sub-sequences and learns a dynamic group of representations according the user evolving preference.
To be specific, we formalize the allocation task as MDP problem and allocate the item in user's sequence behavior into different sub-sequences with a novel action "creating a new action", and form the sub-sequence representations through the sequential recommender model Attention-GRU accordingly.
And 
we conducted experiments to verified the effectiveness of Adasplit on four real datasets with SOTA methods.
However,
the proposed model also exists shortcomings in computing speed,
where we formalize the allocation task in behavior allocator as a MDP problem and it is computing cost and unstable in training.
In the future we will consider how to allocate the user sequence in a more effective way.

\bibliographystyle{ACM-Reference-Format}
\balance
\bibliography{acmart}

\end{document}